\documentclass[twocolumn,prd,nobibnotes,showpacs,showkeys,superscriptaddress]
{revtex4-1}
\usepackage[dvips]{graphicx,graphics,color,epsfig}
\usepackage[colorlinks=true,linkcolor=blue,citecolor=blue,urlcolor=blue,
breaklinks=true]{hyperref}
\usepackage{amsmath,amsfonts,amssymb}
\usepackage{verbatim}
\usepackage{latexsym}
\usepackage{dcolumn}
\usepackage{bm}
\usepackage{natbib}

\begin{document}

\title{Mass characteristics of active and sterile neutrinos\\
in a phenomenological (3+1+2)-model}

\author{\firstname{N.~Yu.}~\surname{Zysina}}
\email{zysina\_ny@nrcki.ru}
\affiliation{National Research Center ``Kurchatov Institute'',
Kurchatov~place~1, 123182~Moscow, Russia}

\author{\firstname{S.~V.}~\surname{Fomichev}}
\email{svfomichev@mail.ru}
\affiliation{National Research Center ``Kurchatov Institute'',
Kurchatov~place~1, 123182~Moscow, Russia}

\author{\firstname{V.~V.}~\surname{Khruschov}}
\email{vkhru@yandex.ru}
\affiliation{National Research Center ``Kurchatov Institute'',
Kurchatov~place~1, 123182~Moscow, Russia}

\begin{abstract}
On the basis of the experimental data allowing existence of the anomalies
going beyond the minimally extended Standard Model with three active neutrinos
with different masses, we consider the generalized model with three active
and three sterile neutrinos, one of which is relatively heavy ((3+1+2)-model).
We study the basic characteristics that are used to describe the massive
active and sterile neutrinos, the methods to determine the absolute scale of
neutrino masses and the neutrino mass estimates based on the available
experimental data. Taking into account the possible contributions of the
sterile neutrinos, the dependences of the neutrino mass characteristics on
the sterile neutrino mass are plotted. The results obtained can be used to
interpret and to predict the results of various neutrino experiments.
\end{abstract}



\maketitle

\section{Introduction}%
\label{Sec1}%
Among the problems of modern neutrino physics the basic ones are the question
about Majorana or Dirac nature of neutrinos, the problem of the number of
different types of neutrinos and the precise measurements of absolute
values of the masses and mixing
parameters of neutrinos. To address these issues, active experimental and
theoretical studies of both the neutrino mass observables that define the
absolute mass scale of neutrino and neutrino oscillation characteristics that
characterize the mixing of neutrinos with different masses are currently
performed. It is expected that a comprehensive solution of the problems of the
nature of neutrinos, their number and values of the mixing parameters will be
given in the Grand Unification Theory (GUT), which does not still exist now in
the generally accepted version \cite{1}. To identify directions for further
development of the existing Standard Model (SM), various phenomenological
models, which use precise experimental data obtained are proposed and studied.
In the framework of these models, the approximate values of the masses and
mixing parameters of neutrinos and the relationships between them can be
obtained, which can play an important role in clarifying the ways to extend
the SM and in further to build the GUT successfully. Of course, for realization
of these aims the crucial role belongs to the data that will become available
as a result of carrying out in the next few years the neutrino experiments
such as PLANK, KATRIN, GERDA, CUORE, BOREXINO, Double CHOOZ, SuperNEMO,
KamLand--Zen, EXO, etc.

Considering the results obtained recently in the neutrino physics, let us to
pay a special attention to the following two: the difference from zero (more
than $5\sigma$) of the reactor mixing angle $\theta_{13}$ \cite{2} and the
experimental evidence of the possible existence of new anomalies for neutrino
and antineutrino fluxes in different processes \cite{3}. The first result is
of great importance for justification of the experimental search for the
{\it CP} violation in the lepton sector \cite{4} (because so far there are no
experimental data to support the lepton {\it CP} violation \cite{1}) and for
definition of the explicit form of the neutrino mixing matrix and neutrino
mass matrix \cite{5}. The second result can be related to the sterile
neutrinos, whose existence is not only beyond the SM, but also beyond the
minimally extended SM with three active neutrinos with different masses
(MESM). Sterile neutrinos, which do not interact with the known conventional
particles can in principle be numerous. The most popular models
for sterile neutrinos now are the phenomenological models with one or two
sterile neutrinos. These are so-called $(3+1)$- and $(3+2)$-models \cite{6,7}.
However, if we take into account a possible left-right symmetry of the weak
interactions and identify sterile neutrinos with the right singlet neutrinos
with respect to the $SU(2)_L$, then the number of sterile neutrinos should be
three, and so we come to the less popular $(3+3)$-models. They are less
popular for the two reasons: to explain the experimentally observed anomalies
it is sufficient, as it is now considered, to introduce only two sterile
neutrinos, and, in addition, recent data of the cosmological observations of
the fluctuations of the cosmic microwave background barely allow us to add one
additional type of neutrino, but to add extra one, not to speak about two new
types of neutrino - this is a serious problem. On the other hand, as was noted
above, the right-left symmetric models should operate with three sterile
neutrinos \cite{8,9,10}. Models with three sterile neutrinos have been
considered in Ref.~\cite{11}, while in Ref.~\cite{12} a model with a light
sterile neutrino with a mass of the order of the mass of active neutrinos has
been proposed. In this paper we consider the phenomenological $(3+1+2)$-model
with one relatively heavy sterile neutrino and two light sterile neutrinos
\cite{8,9}. In this model it is easy if necessary to reduce the number of
sterile neutrinos to avoid the possible contradictions with the data of
cosmological observations \cite{13,14}. However, there is still a possibility
to keep the number of different sterile neutrinos equal to three, but for this
it is necessary to reduce the link between the active and the sterile
neutrinos and, for example, to decline the condition of thermal equilibrium
between the sterile neutrinos and other relativistic particles in the early
stages of the Universe \cite{15,16,17}.

In this paper, a $(3+1+2)$-model is used for the possible solution of the two
problems in the neutrino physics, namely, how to include the sterile neutrinos
in a generalized model of weak interactions to explain the new detected
anomalies in the neutrinos and antineutrinos fluxes in the various processes,
and what a value takes the mass scale of the active and sterile neutrinos? As
is known, the results of experiments involving neutrino oscillations can
determine the values of differences of masses in square of these particles and
not the mass values themselves that leads to the problem of neutrino mass
hierarchy and the problem of their mass scale. These problems have become more
complex due to the possible existence of sterile neutrinos with masses of
the order of $1\,{\rm eV}$. An extensive literature is devoted to
consideration of these questions; note only some of the recent reviews
\cite{7,18,19}.

The paper is organized as follows. In Section~\ref{Sec2}, the main
characteristics, which are used to describe the Dirac and Majorana massive
neutrinos, as well as the available experimental data for these
characteristics are given. The experimental evidences for the existence of
anomalies that go beyond the Standard Model with three active neutrinos are
discussed in Section~\ref{Sec3}. The main statements of the phenomenological
$(3+1+2)$-model of neutrino are described in Section~\ref{Sec4}.
In Section~\ref{Sec5}, we present the phenomenological estimates of the masses
of both active and sterile neutrinos. Particular attention is paid in
Section~\ref{Sec6} to the methods of determination of the absolute scale of
neutrino masses with the help of the mass observables and to the experimental
limitations on possible values of the neutrino mass observables. In this
section we also present the results of numerical calculations of the neutrino
mass characteristics in the form of plots versus the possible values of the
minimal neutrino mass, which take into account the contributions associated
with the sterile neutrinos. In final Section~\ref{Sec7} we discuss the results
of the paper, which can be used to interpret and to predict the results of
various experiments on determination of the neutrino mass scale and on
searching for the effects associated with sterile neutrinos.

\vspace{5mm}%
\section{Oscillation characteristics of the Dirac and Majorana massive
neutrinos}%
\label{Sec2}%
As is known, the oscillations of solar, atmospheric, reactor and accelerator
neutrinos can be explained by mixing of the neutrino states. This means that
the flavor neutrino states are a mixture of at least three massive neutrino
states and vice versa. Neutrino mixing is described by
the Pontecorvo--Maki--Nakagawa--Sakata matrix $U_{PMNS}\equiv U=VP$:
\begin{equation}
\psi^{\alpha}_L=U^{\alpha}_i\psi^i_L\,,
\label{eq1}
\end{equation}
where $\psi^{\alpha}_L$ and $\psi^{i}_L$ are the left chiral fields of flavor
and massive neutrinos, respectively, with $\alpha={e,\mu,\tau}$ and
$i={1,2,3}$. For the three active neutrinos, matrix $V$ can be written in the
standard parametrization as \cite{1}
\begin{widetext}
\begin{equation}
V=\left(\begin{array}{lcr}
c_{12}c_{13}&s_{12}c_{13}&s_{13}e^{-i\delta_{CP}}\\
-s_{12}c_{23}-c_{12}s_{23}s_{13}e^{i\delta_{CP}}
&c_{12}c_{23}-s_{12}s_{23}s_{13}e^{i\delta_{CP}}&s_{23}c_{13}\\
s_{12}s_{23}-c_{12}c_{23}s_{13}e^{i\delta_{CP}}
&-c_{12}s_{23}-s_{12}c_{23}s_{13}e^{i\delta_{CP}}&c_{23}c_{13}
\end{array}\right),
\label{eq2}
\end{equation}
\end{widetext}
where $c_{ij}\equiv\cos\theta_{ij}$, $s_{ij}\equiv\sin\theta_{ij}$,
$\delta_{CP}$ is the phase associated with the Dirac {\it CP} violation in the
lepton sector, $P={\rm diag}\{1,e^{i\alpha_{CP}},e^{i\beta_{CP}}\}$, and
$\alpha_{CP}$ and $\beta_{CP}$ are the phases related to the Majorana
{\it CP} violation.

In a general case, the unitary $n\times n$-matrix is determined by $n^2$ real
parameters, for which we may choose $n(n-1)/2$ angles and $n(n+1)/2$ phases.
Taking into account the structure of the electroweak SM Lagrangian, which
comprises the currents composed of the fields of quarks, charged leptons and
neutrinos, in the case of the Dirac neutrino fields it is possible to
eliminate the $2n-1$ phases. In the case when the neutrino fields are of the
Majorana type, one can exclude only $n$ phases associated with the Dirac
charged leptons. In view of this, $n\times n$-matrix $U$ depending on the
nature of neutrinos is determined either by $n(n-1)/2$ angles and
$(n-1)(n-2)/2$ phases if neutrinos are of the Dirac nature, or by $n(n-1)/2$
angles and $n(n-1)/2$ phases if neutrinos are the Majorana particles
\cite{20}. Thus, to set the mixing matrix $U_{PMNS}$ in the case of three
Dirac neutrinos it is necessary to determine the three mixing angles
$\theta_{12}$, $\theta_{13}$, $\theta_{23}$ and one mixing phase
({\it CP}-phase $\delta_{CP}$), while for three Majorana neutrinos one should
determine the same three mixing angles ($\theta_{12}$, $\theta_{13}$,
$\theta_{23}$) and three $CP$-phases $\delta_{CP}$, $\alpha_{CP}$ and
$\beta_{CP}$.

Because the {\it CP} violation in the lepton sector is not yet experimentally
observed \cite{1}, in further we will consider the elements of the neutrino
mixing matrix as real. Generally, when the number of different types of
neutrinos is $3+N$, the neutrino mass matrix $M_{\nu}$ can be determined with
the help of the generalized mixing matrix $\tilde{U}$ of rank
$(3+N)\!\times\!(3+N)$ as follows
\begin{equation}
M_{\nu}=\tilde{U}M^d\tilde{U}^T\,,
\label{eq3}
\end{equation}
where $M^d={\rm diag}\{m_1,m_2,m_3,\dots,m_{3 + N}\}$ and
$m_i$ ($i=1,2,3,\dots,3+N$) are the masses of the neutrinos. Thus, the matrix
elements $M_{\nu,ij}$ depend on the values of the masses and mixing parameters
of the neutrinos.

As is known, with using only the oscillation experiments with atmospheric,
solar, reactor and accelerator neutrinos it is not possible to determine the
absolute values of the neutrino masses, as well as their Dirac or Majorana
nature. The obtained experimental results related to neutrino oscillations
indicate a violation of the laws of conservation of lepton numbers $L_e$,
$L_{\mu}$ and $L_{\tau}$ and, in addition, the existence of at least two
non-zero and different neutrino masses. The latter is a consequence of the
non-zero values of two oscillation parameters $\Delta m_{12}^2$ and
$\Delta m_{13}^2$ (where $\Delta m_{ij}^2=m_i^2-m_j^2$). Three sets of
experimental data are sensitive to the absolute scale of neutrino masses,
namely, the experimental data on beta decay, the experimental data on
neutrinoless double beta decay and the experimental data obtained as a result
of cosmological observations of the structure of the early universe at large
distances. With the help of appropriate data for each of these sets,
respectively, one of three neutrino mass observables, namely, $m_{\beta}$,
or $m_{\beta\beta}$, or $m_{\Sigma}$, which are defined below can be measured.

Here we display the values of the mixing angles and neutrino mass-squared
differences obtained from the overall analysis of the latest high-precision
measurements of the oscillation parameters, which determine three-flavor
oscillations of light active neutrinos with the standard deviations at the
level of $1\sigma$ \cite{2}:
\begin{subequations}
\begin{align}
&\sin^2\theta_{12}=0.307^{+0.018}_{-0.016}\,,\label{eq4a}\\
&\nonumber\\
&\sin^2\theta_{23}=\left\{\begin{array}{lr}NH:&0.386^{+0.024}_{-0.021}\\
IH:&0.392^{+0.039}_{-0.022}
\end{array}\right.\!,\label{eq4b}\\
&\nonumber\\
&\sin^2\theta_{13}=\left\{\begin{array}{lr}NH:&0.0241^{+0.0025}_{-0.0025}\\
IH:&0.0244^{+0.0023}_{-0.0025} \end{array}\right.\!,\label{eq4c}\\
&\nonumber\\
&\Delta m_{21}^2/10^{-5}{\rm eV}^2=7.54^{+0.26}_{-0.22}\,,\label{eq4d}\\
&\nonumber\\
&\Delta m_{31}^2/10^{-3}{\rm eV}^2=
\left\{\begin{array}{lr}NH:&2.47^{+0.06}_{-0.10}\\
IH:&-2.46^{+0.07}_{-0.11} \end{array}\right.\!.
\label{eq4e}
\end{align}
\label{eq4}
\end{subequations}
Note that {\it CP}-phases $\alpha_{CP}$, $\beta_{CP}$ and $\delta_{CP}$ are
not currently known, as well as the scale of neutrino masses. Since we know
only the absolute value of $\Delta m_{31}^2$, the absolute values of the
neutrino masses can be ordered by two ways, namely: a) $m_1<m_2<m_3$ and b)
$m_3<m_1<m_2$, i.e., it can be realized, as they say, either the normal
hierarchy (NH, in the case a) or the inverse hierarchy (IH, in the case b) of
the neutrino mass spectrum.

\section{Anomalies of neutrino fluxes from different sources}%
\label{Sec3}%
The results of a number of neutrino experiments at small distances (more
exactly, at distances $L$, at which the numerical value of the parameter
$L\Delta m^2/E$ is of the order of unity, with $E$ the neutrino energy)
indicate the possible existence of light sterile neutrinos with masses of the
order of $1\,{\rm eV}$ \cite{3}. Discovery of such neutrinos would be a
fundamental contribution to the physics of weak interactions. Sterile
neutrinos are involved to explain the experimental data, which can not be
explained in terms of common three-flavor model of mixing of massive neutrinos.
Firstly, it is the so-called LSND/MiniBooNE anomalies \cite{21,22}. Besides,
it has recently been carried out refined calculations of the spectra of
reactor antineutrinos \cite{23}, which result in higher calculated values of
the fluxes of these particles. That is, the obtained experimental data
indicate a deficiency of antineutrino fluxes under the measurements at
distances from the source less than $100\,{\rm m}$. Such distances from the
source are called as small, and the experiments at such distances are called
as SBL (shot baseline) experiments. Possible deficiency of reactor
antineutrinos at low distances is called as the reactor anomaly \cite{24}.
Similar anomalies were observed in calibration measurements for experiments
SAGE and GALLEX. Such anomalies are commonly called as calibration or gallium
ones \cite{25,26,27}. From this it follows that characteristic values of
$\Delta m^2$ for sterile neutrinos are about $1\,{\rm eV}^2$. Thus, currently
available three types of anomalies for neutrino fluxes (LSND/MiniBooNE,
reactor, gallic) can be interpreted as evidence on the level of abour
$3\sigma$ for the existence of sufficiently heavy sterile neutrinos. However,
the additional verification of these anomalies are required.

Below we consider the phenomenological $(3+1+2)$-model of the neutrino
\cite{8,9,KF2013arx}, which can be used for incorporation of sterile neutrinos
in the formalism of the theory of weak interactions. This model contains three
active light neutrinos and three sterile neutrinos, and one sterile neutrino
is relatively heavy, while the other two are light sterile neutrinos. Assuming
all the cosmological limitations on the total number of neutrinos
\cite{13,14}, the number of sterile neutrinos can be reduced in this model,
if necessary. However, such a reduction should be stipulated by weighty
reasons. Since the cosmological constraints on the total number of neutrinos
obtained on the basis of observational data are model-dependent, more studies
are needed concerning the development of the cosmological models that allow,
at joint variation of other cosmological model parameters, the insertion of
one to three new types of neutrinos (and neutrinos hardly interacting with
the particles of the SM).

\section{The basic points of the phenomenological (3+1+2)-model of the
neutrino}%
\label{Sec4}%
Consider the formalism of neutrino mixing for a generalized model of the weak
interactions, in which there are three active neutrinos and three sterile
neutrinos, i.e., the $(3+3)$-model. This model can be easy reduced to
$(3+2)$- or even $(3+1)$-model, with reducing the number of sterile neutrinos.
Different types of sterile neutrinos will be distinguished by indices $x$, $y$
and $z$, and additional massive states will be distinguished by indexes $1'$,
$2'$ and $3'$. The set of indices $x$, $y$ and $z$ will be denoted by common
symbol $a$, and the set of indices $1'$, $2'$ and $3'$ will be denoted by
symbol $i'$. Then the mixing matrix $\tilde{U}$ of rank $6\!\times\!6$ can be
represented in a block form with use of four $3\!\times\!3$ matrices $S$, $T$,
$V$ and $W$, so that
\begin{equation}
\begin{pmatrix}
\nu_{\alpha} \\ \nu_a
\end{pmatrix}=\tilde{U}
\begin{pmatrix}
\nu_{i} \\ \nu_{i'}
\end{pmatrix}\equiv
\begin{pmatrix}
S & T \\ V & W
\end{pmatrix}
\begin{pmatrix}
\nu_{i} \\ \nu_{i'}
\end{pmatrix}.
\label{eq5}
\end{equation}
Neutrino masses will be given by a set $\{m\}=\{m_i,m_{i'}\}$, in which
$\{m_i\}$ will be arranged in the normal order as $\{m_1,m_2,m_3\}$, while
$\{m_{i'}\}$ will be arranged in the reverse order as
$\{m_{3'},m_{2'},m_{1'}\}$. In the current paper, the unitary matrix
$\tilde{U}$ will not be considered in the most general form, but we restrict
ourselves only by special cases involving additional assumptions
\cite{KF2013arx}. For example,
on the basis of available data of laboratory and cosmological observations it
can be assumed that the mixing between the active and the sterile neutrinos is
small, and, moreover, as the basis of the massive sterile states we choose the
states, for which the matrix $W$ is the unit matrix. With keeping in the
unitarity condition only the first-order terms, which are contained in small
$3\!\times\!3$ matrices $b$ and $\Delta U_{PMNS}$, we can write the matrices
$S$, $T$ and $V$ as
\begin{subequations}
\begin{align}
&S=U_{PMNS}+\Delta U_{PMNS}\,,\label{eq6a}\\
&\nonumber\\
&T=b, \quad V=-b^TU_{PMNS}\,.\label{eq6b}
\end{align}
\label{eq6}
\end{subequations}
The contributions from the matrix elements of $\Delta U_{PMNS}$ in fact are
already taken into consideration by using the experimental uncertainties of
the matrix $U_{PMNS}$ itself. Taking into account that in explaining the
discovered anomalies in the neutrino fluxes the most attention should be paid
to the adjustments of the neutrino flavor states $\nu_e$ and $\nu_{\mu}$,
which are associated with the most massive sterile states, the matrix $b$ can
be chosen in the following manner: in the IH-case
\begin{equation}
b_{IH}=\begin{pmatrix}
\gamma & \delta & \delta \\
\beta & \delta & \delta \\
\alpha & 0 & 0
\end{pmatrix},
\label{eq7}
\end{equation}
whereas in the NH-case
\begin{equation}
b_{NH}=\begin{pmatrix}
0 & 0 & \alpha \\
\delta & \delta & \beta \\
\delta & \delta & \gamma
\end{pmatrix},
\label{eq8}
\end{equation}
where on the basis of the available experimental constraints \cite{7,18,19,28}
the parameters $\alpha$, $\beta$, $\gamma$ and $\delta$ are small quantities
lesser than $0.2$ in absolute value. In addition, marking out among the mass
of sterile neutrinos the greatest one we will consider that the masses of
other sterile neutrinos are smaller by at least one order of magnitude. This
choice of the masses is consistent with the estimates of the masses of sterile
neutrinos given in the next section. Therefore, taking into account the
accepted mass distribution of the sterile neutrinos, the $(3+3)$-model in
this case can be called as the $(3+1+2)$-model of active and sterile neutrinos.

\vspace{3mm}%
\section{Phenomenological estimates of neutrino masses}%
\label{Sec5}%
\vspace{2mm}%
As is known, the problem of strongly differing masses of the fundamental
fermions remains still unsolved. In the SM these masses arise from the Yukawa
couplings between the fields of the fundamental fermions and the Higgs
field. However, the values of neutrino masses are so small that probably the
mechanism of their formation is mainly due to the Majorana nature of
neutrinos. In this case, the main task is to determine the special mechanism
of forming the masses of the Majorana neutrinos. In the absence of a
satisfactory theory of this phenomenon the problem was considered at the
phenomenological level by many authors \cite{8,9,29,30,31,32}. We will use
the results of Refs.~\cite{8,KF2013arx}, in which the following estimates
of the absolute values of the masses of the light active neutrinos
$m_i$ ($i=1,2,3$) were obtained. For the NH-case, they are [in {\rm eV}]
\begin{subequations}
\begin{equation}
m_1\approx 0.0015, \quad m_2\approx 0.0088, \quad m_3\approx 0.0497\,,
\label{eq9a}
\end{equation}
and for the IH-case they are
\begin{equation}
m_1\approx 0.0496, \quad m_2\approx 0.0504, \quad m_3\approx 0.0020\,.
\label{eq9b}
\end{equation}
\label{eq9}
\end{subequations}
Mass estimates for the right-handed neutrinos $M_i$ for these two cases result
in the following values, respectively:
\begin{subequations}
\begin{align}
&M_1\approx\Lambda_1, \quad  M_2\approx 0.002, \quad M_3\approx 0.002\,,
\label{eq10a}\\
&\nonumber\\
&M_1\approx 0.002, \quad  M_2\approx 0.002, \quad M_3\approx\Lambda_3\,,
\label{eq10b}
\end{align}
\label{eq10}
\end{subequations}
where $\Lambda_{1,3}$ are the free parameters of the order of $1\,{\rm eV}$.
As was noted above, it is possible to compare the right neutrinos $\nu_{Ri}$
($i=1,2,3$) to the sterile neutrinos $\nu_{i'}$
($i'=1^{\prime},2^{\prime},3^{\prime}$), i.e., to equate masses $M_i$ to
masses $m_{i'}$. It results in separation of the three sterile neutrinos
with respect to masses into two light and one relatively heavy neutrino. Light
right sterile neutrinos can be combined with light left neutrinos and form
quasi-Dirac neutrinos. The case with three light active neutrinos, one heavier
sterile neutrino and two light sterile neutrinos was called in
Refs.~\cite{8,9} as the $(3+2+1)$-model. This model can be reduced, if
necessary, to $(3+1+1)$- and even $(3+1)$-model, eliminating the light
right-handed neutrinos, which are singlets of the gauge group
$SU(2)_L\times U(1)_Y$ of the weak and electromagnetic interactions.

The given above values of neutrino masses can be used, for example, for
estimates of the values of the anomalous magnetic moments of neutrinos
\cite{33}. As a part of the MESM with non-zero neutrino masses, the neutrino
magnetic moments are due to the radiative corrections. In general, the
magnitudes of the magnetic moments, which are proportional to the masses of
neutrinos form a matrix, the size of which depends on the number of neutrino
flavours, and for the Majorana neutrinos the diagonal magnetic moments equal
to zero. The diagonal Dirac moments in the one-loop approximation are as
follows \cite{34,35}
\begin{equation}
\mu_i =(3G_Fm_em_i/4\pi^2\sqrt{2})\mu_B \quad (i=1,2,3)\,,
\label{eq11}
\end{equation}
where $\mu_B$ is Bohr magneton, $m_e$ is mass of the electron, $G_F$ is Fermi
constant of the weak interactions, i.e.,
$\mu_i/\mu_B\approx 3.203\times 10^{-19}\times m_i[{\rm eV}]$.

Using estimates of the neutrino masses given in
Eqs.~(\ref{eq9})--(\ref{eq10}), when the right-handed neutrinos are identified
with light sterile neutrinos, we find that the massive Dirac neutrinos must
have magnetic moments of the order of $10^{-20}\mu_B$. In the MESM, the
off-diagonal magnetic moments of neutrino are suppressed by several orders of
magnitude as compared with the diagonal magnetic moments. However, there are
some models that go beyond the MESM, in which the off-diagonal Majorana
neutrino magnetic moments can be much larger. The most optimistic in this
respect are extended models with restored left-right gauge symmetry. For
example, in such models for the off-diagonal neutrino magnetic moments
$\mu_{e\tau}$ and $\mu_{\mu\tau}$ the estimates are obtained on the level of
about $(10^{-13}\div 10^{-14})\mu_B$, and these values depend on the mixing
angle between the left and right gauge $W$-bosons \cite{36}. The use of such
models is most justified in interpreting the sterile neutrinos as a
right-handed neutrinos, which is adopted in this paper.

The above estimates of relatively large magnetic moments of Majorana neutrinos
in the framework of the right-left symmetric models, if they are realized,
will affect, for example, the dynamics of flavour conversion of neutrinos
inside the neutrino-sphere formed in supernova explosions \cite{37}. Since
the magnitude of the Dirac magnetic moments in MESM is about $10^{-20}\mu_B$,
it can be concluded that the detection of the magnetic moments of the order
of $10^{-14}\mu_B$ would be evidence in favour of that the neutrinos are the
Majorana particles. This fact would be of great importance for creating a
realistic GUT and for searching the processes, in which there is a
violation of the conservation law of total lepton number, for example, for
searching the neutrinoless double beta decay.

\section{Neutrino mass observables with allowance for the sterile neutrino
contributions}%
\label{Sec6}%

\begin{figure*}
\includegraphics[width=0.95\textwidth]{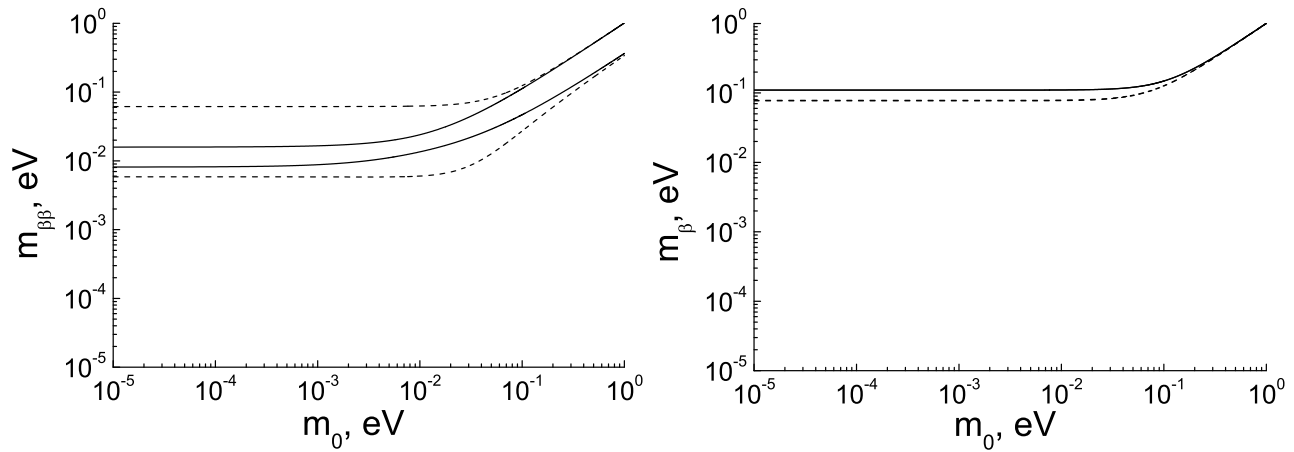}
\caption{The ranges of variation of effective neutrino masses $m_{\beta\beta}$
(left panel) and $m_{\beta}$ (right panel) as a function of the smallest
neutrino mass $m_0$. The ranges between the solid lines correspond to the
NH-case, while the ranges between the dashed lines correspond to the IH-case
(in the case of $m_{\beta}$, they are extremely narrow and degenerate into
lines). $\alpha=0.1$, $\gamma=0.1$, $\delta=0.1$, and $m_*=0.3\,{\rm eV}$,
where $m_*=m_{1'}$ for the NH-case and $m_*=m_{3'}$ for the IH-case, and the
values of the masses $m_{1'}$ and $m_{2'}$ are equal in the absolute value to
$0.002\,{\rm eV}$.}\label{Fig1}
\end{figure*}
\begin{figure*}
\includegraphics[width=0.95\textwidth]{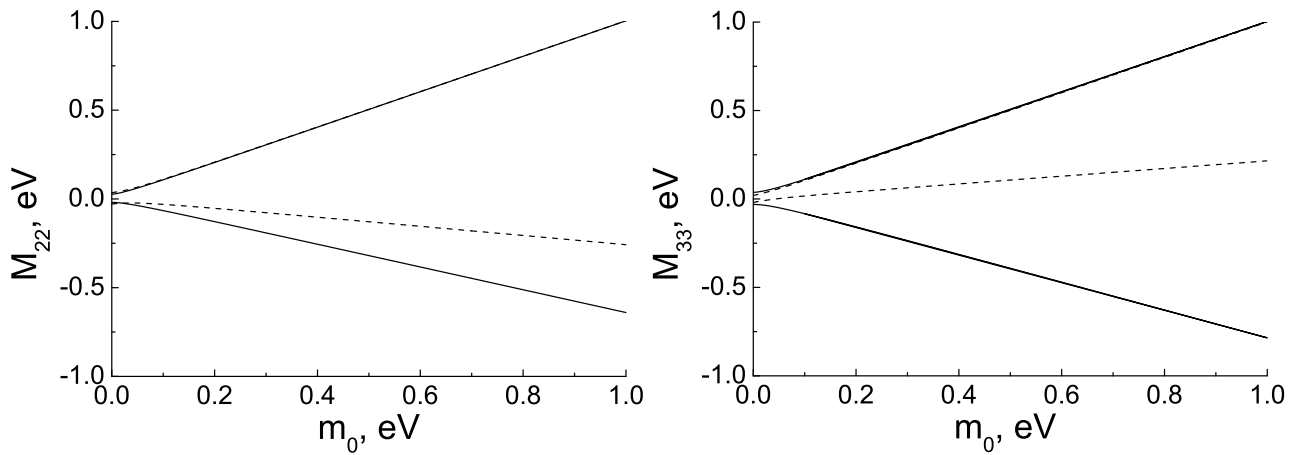}
\caption{The ranges of variation of the matrix elements $M_{22}$ and $M_{33}$
of the neutrino mass matrix. The ranges between the solid lines correspond to
the NH-case, while the ranges between the dashed lines correspond to the
IH-case (with the virtually same lines for {\it maximal} values in both
cases). The parameters are the same as in Fig.~\ref{Fig1}.}\label{Fig2}
\end{figure*}
\begin{figure*}
\includegraphics[width=0.97\textwidth]{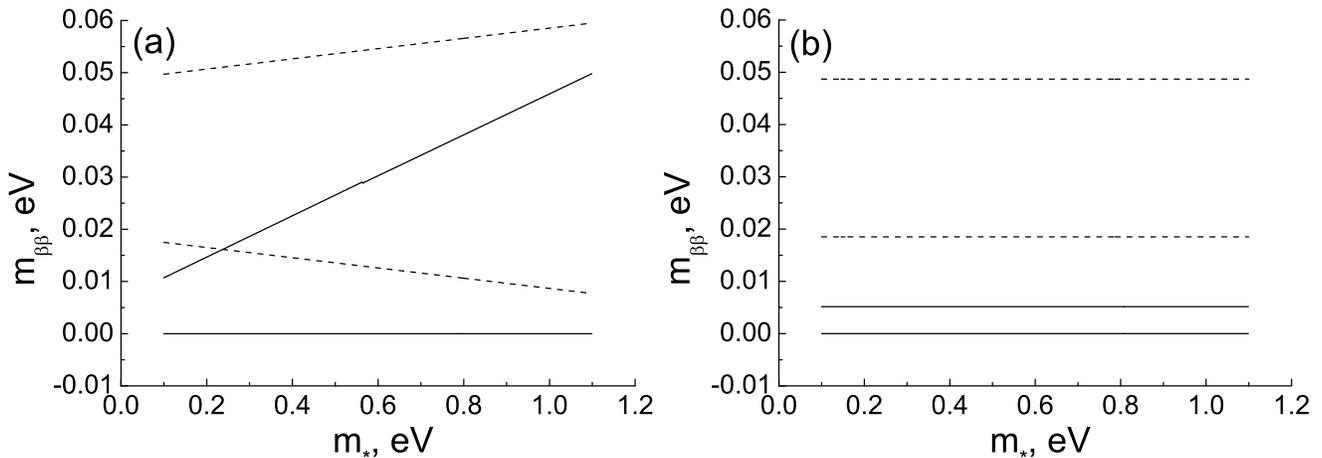}
\caption{The ranges of variation of the effective mass of neutrinos
$m_{\beta\beta}$, as a function of the greatest mass of sterile neutrinos
$m_*$. The ranges between the solid lines correspond to the NH-case, while the
ranges between the dashed lines correspond to the IH-case. On panel (a), the
variation of the parameters of sterile neutrinos was performed in the range
of $|\alpha|<0.2$, $|\gamma|<0.2$ and $|\delta|<0.2$. For comparison, panel
(b) exhibits the case of $\alpha=\gamma=\delta=0$. For the NH-case,
$|m_1|=0.002$, $|m_2|=0.0087$, $|m_3|= 0.0497$, $|m_{2'}|=0.002$ and
$|m_{1'}|=0.002$, while for the IH-case $|m_1|=0.0496$, $|m_2|=0.05$,
$|m_3|=0.002$, $|m_{2'}|=0.002$ and $|m_{1'}|=0.002$.}\label{Fig3}
\end{figure*}
\begin{figure*}
\includegraphics[width=0.97\textwidth]{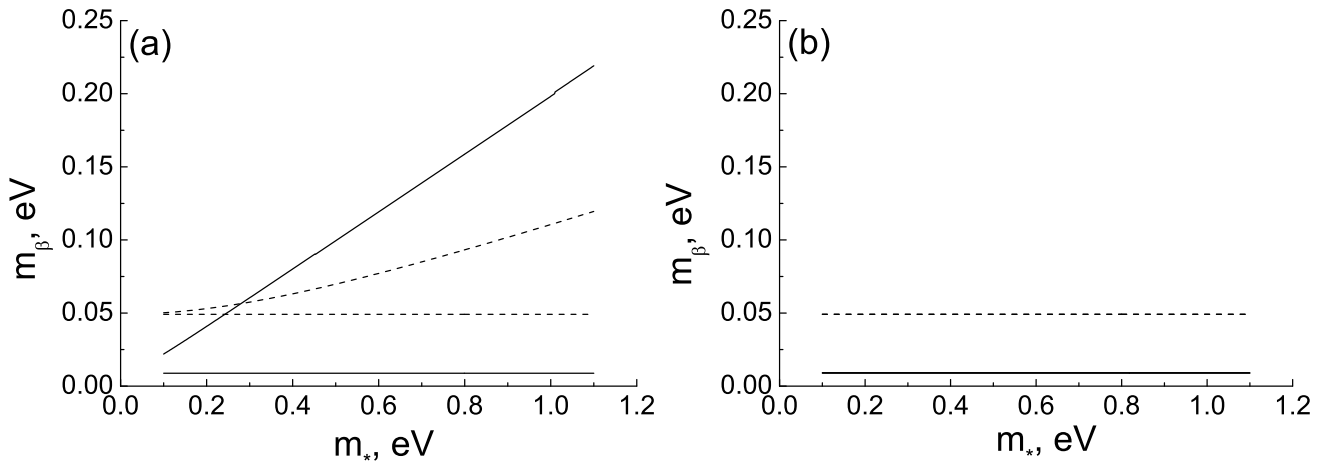}
\caption{The ranges of variation of the effective mass of neutrinos
$m_{\beta}$. The ranges between the solid lines correspond to the NH-case,
while the ranges between the dashed lines correspond to the IH-case. On panel
(a), the variation of the parameters of sterile neutrinos was performed in
the range of $|\alpha|<0.2$, $|\gamma|<0.2$ and $|\delta|<0.2$. For
comparison, panel (b) exhibits the case of $\alpha=\gamma=\delta=0$ (note that
here the lines for minimal and maximal values practically coincide in both
(NH and IH) cases, so the ranges degenerate into lines). The other parameters
are the same as in Fig.~\ref{Fig3}.}\label{Fig4}
\end{figure*}
\begin{figure*}
\includegraphics[width=0.94\textwidth]{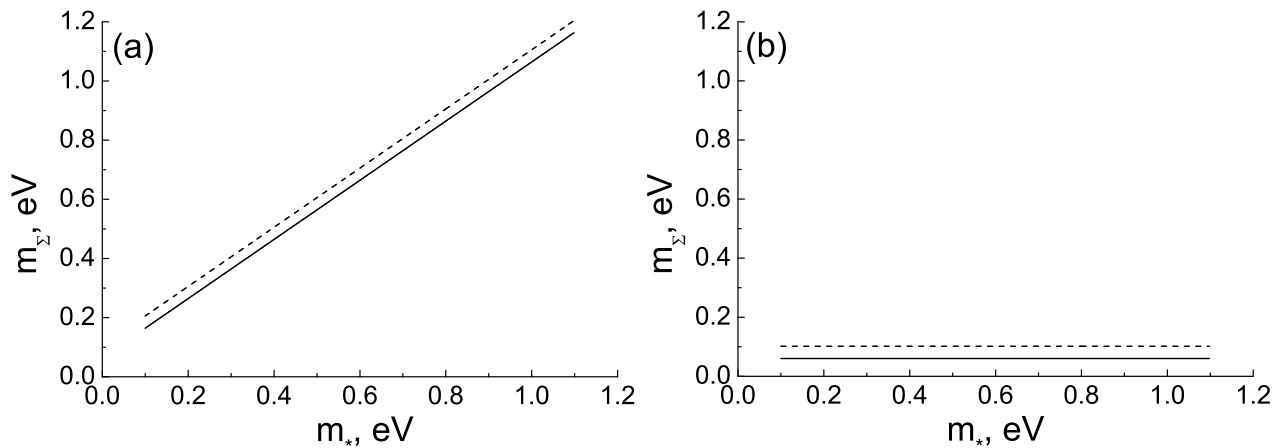}
\caption{The ranges of variation of the cosmological effective mass of
neutrinos $m_{\Sigma}$ (and here the lines for minimal and maximal values
practically coincide in both (NH and IH) cases, so the ranges degenerate into
lines). The like-lines ranges between the solid lines correspond to the
NH-case, while ones between the dashed lines correspond to the IH-case. On
panel (a), the variation of the parameters of sterile neutrinos was performed
in the range of $|\alpha|<0.2$, $|\gamma|<0.2$ and $|\delta|<0.2$. For
comparison, panel (b) exhibits the case of $\alpha=\gamma=\delta=0$. The other
parameters are the same as in Fig.~\ref{Fig3}.}\label{Fig5}
\end{figure*}
\begin{figure*}
\includegraphics[width=0.94\textwidth]{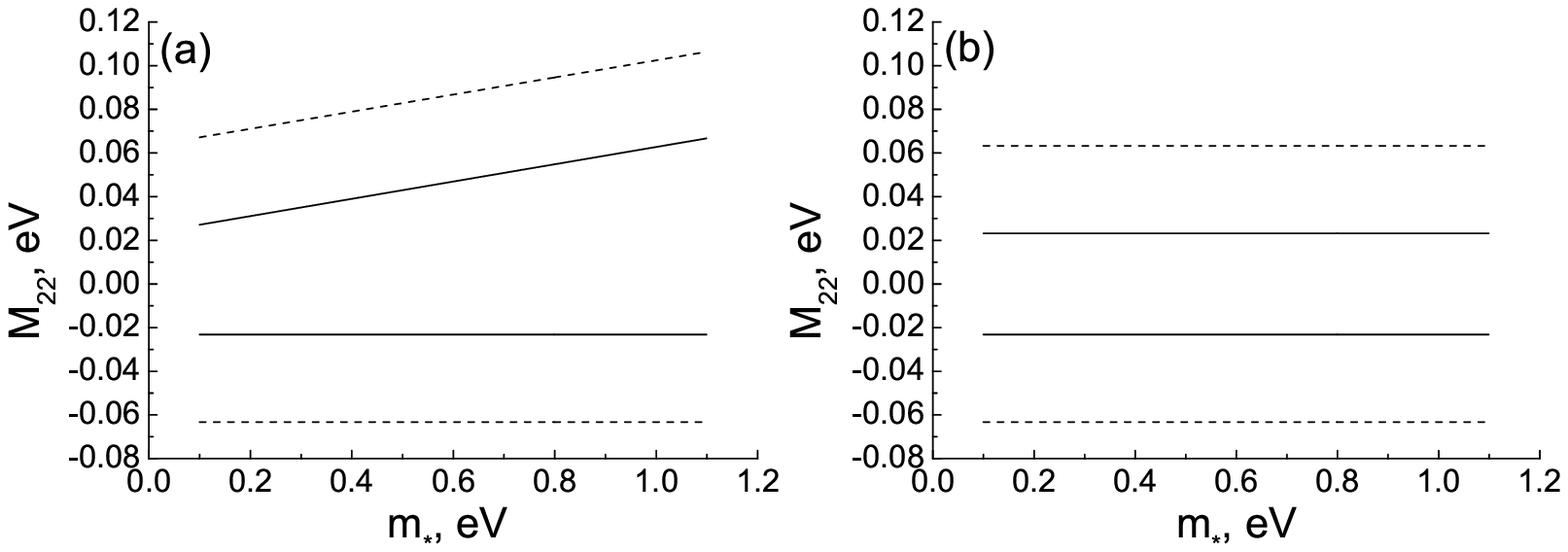}
\caption{The ranges of variation of the matrix element $M_{22}$. The ranges
between the solid lines correspond to the NH-case, while ones between the
dashed lines correspond to the IH-case. On panel (a), the variation of the
parameters of sterile neutrinos was performed in the range of $|\alpha|<0.2$,
$|\gamma|<0.2$ and $|\delta|<0.2$. For comparison, panel (b) exhibits the case
of $\alpha=\gamma=\delta=0$. The other parameters are the same as in
Fig.~\ref{Fig3}.}\label{Fig6}
\end{figure*}
\begin{figure*}
\includegraphics[width=0.95\textwidth]{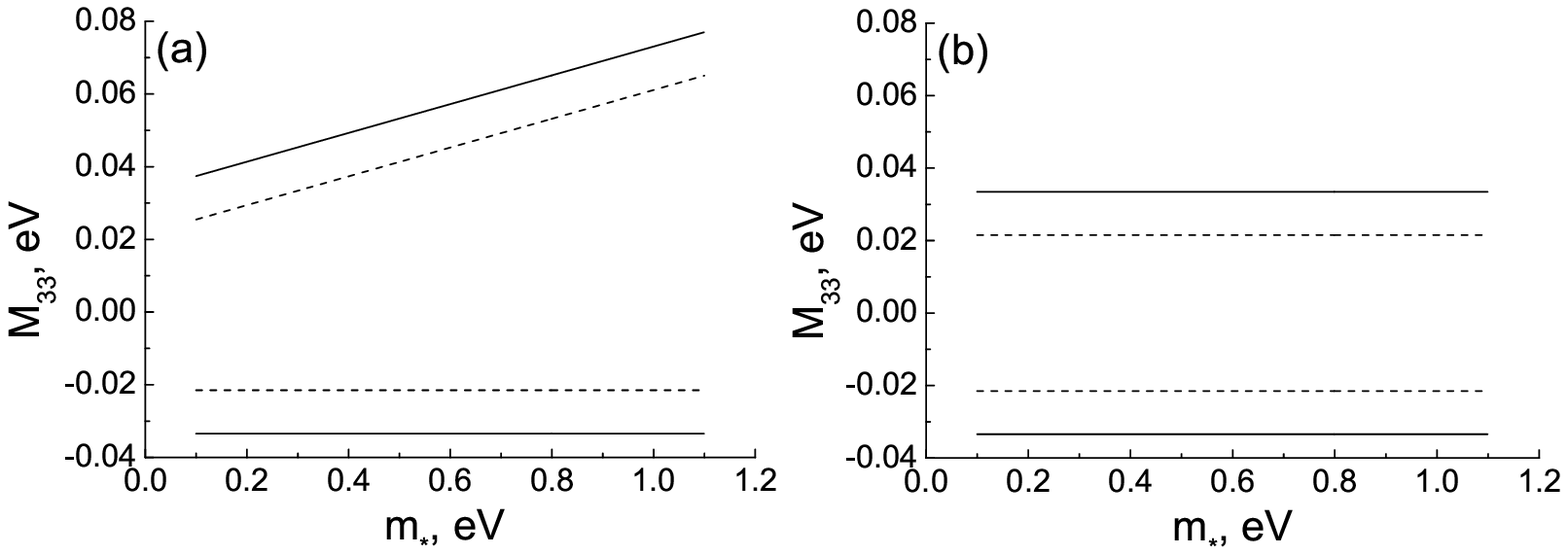}
\caption{The ranges of variation of the matrix element $M_{33}$. The ranges
between the solid lines correspond to the NH-case, while ones between the
dashed lines correspond to the IH-case. On panel (a), the variation of the
parameters of sterile neutrinos was performed in the range of $|\alpha|<0.2$,
$|\gamma|<0.2$ and $|\delta|<0.2$. For comparison, panel (b) exhibits the case
of $\alpha=\gamma=\delta=0$. The other parameters are the same as in
Fig.~\ref{Fig3}.}\label{Fig7}
\end{figure*}

In order to determine the absolute scale of neutrino masses, it is necessary
to determine experimentally at least one of the following values of the
neutrino mass observables, namely, the cosmological sum of the neutrino masses
$m_{\Sigma}$, the $\beta$-decay neutrino mass $m_{\beta}$ or the effective
(double $\beta$-decay) neutrino mass $m_{\beta\beta}$:
\begin{subequations}
\begin{align}
&m_{\Sigma}=\sum_{i=1}^6\nolimits|m_i|\,,\label{eq12a}\\
&\nonumber\\
&m_{\beta}=\Big(\sum_{i=1}^6\nolimits|U_{ei}|^2m_i^2\Big)^{1/2}\,,
\label{eq12b}\\
&\nonumber\\
&m_{\beta\beta}=\Big|\sum_{i=1}^6\nolimits U_{ei}^2m_i\Big|\,.\label{eq12c}
\end{align}
\label{eq12}
\end{subequations}
In general, the neutrino mass observables $m_{\Sigma}$, $m_{\beta}$ and
$m_{\beta\beta}$, the neutrino masses $m_i$ and $m_{i'}$ and the elements of
the neutrino mass matrix $M_{ij}$ can be called as neutrino mass
characteristics. To the moment, only the experimental upper limits on the
values of the neutrino mass observables are obtained, and they are
$m_{\Sigma}<0.93\,{\rm eV}$ \cite{13,14}
and $m_{\beta\beta}<0.25\,{\rm eV}$ \cite{39}. The results of the
experiments in Troitsk and Mainz on the measurement of the electron energy
spectrum in the $\beta$-decay of tritium give the limitations on the
$\beta$-decay antineutrino mass $m_{\beta}$ ($m_{\beta}<2.05\,{\rm eV}$
\cite{Lobashev}, $m_{\beta}<2.2\,{\rm eV}$ \cite{Lobashev,38}), which is
expected to improve to $0.2\,{\rm eV}$ in the planned KATRIN experiment
\cite{38}.

Right-handed neutrinos, if they exist, are definitely sterile and are the
candidates for the dark matter particles, and at that, it may be several
different types of dark matter particles with different masses. In some
approximation, the fields $\nu_R$ of right-handed neutrinos may be included
along with the corresponding fields $\nu_L$ in the Dirac mass term in the
Lagrangian, so the right-handed neutrinos will be approximately degenerate in
mass with $\nu_L$, and then, such neutrinos are light particles with masses
less than, at least, $0.3\,{\rm eV}$. However, there are still other
possibilities for the values of the masses of right-handed neutrinos. One of
them is that the masses of $\nu_R$ will be much larger than the masses of
$\nu_L$, but still of the order of $1\,{\rm eV}$. For example, their masses
can be between $0.3\,{\rm eV}$ up to $3\,{\rm eV}$. These neutrinos can be
called as heavy sterile neutrinos. Neutrinos with masses from $3\,{\rm eV}$
up to $3\,{\rm GeV}$ will be called as extra-heavy, and they are super-heavy
if their masses are more than $3\,{\rm GeV}$. Existence of super-heavy
right-handed neutrinos can be used as an explanation for the large quantities
of the hidden mass of the Universe, and also to explain the extremely small
masses of left-handed neutrinos. Moreover, with the help of super-heavy
right-handed neutrinos one can explain the observed baryon asymmetry of the
Universe \cite{40}. In this paper we consider the heavy sterile neutrinos and
also light sterile neutrinos, which, together with the active neutrinos,
in principle, can form quasi-Dirac neutrinos and which are attractive from
a phenomenological point of view \cite{8,9}.

Consider the neutrino mass matrix $M_{ij}$ given by the equation (\ref{eq3}).
The upper diagonal element of this matrix, i.e., $M_{11}$, enters a formula
for the probability of neutrinoless double beta decay of the nucleus,
$(A,Z)\to(A,Z+2)+2e$, if the decay occurs with the participation of the light
Majorana active neutrino. In this case, the absolute value of $M_{11}$
coincides with $m_{\beta\beta}$ of the equation (\ref{eq12c}), i.e., with the
effective mass of the neutrino. Half-life time $T_{1/2}^{0\nu2\beta}$ of the
neutrinoless double beta decay is
inversely proportional to $m_{\beta\beta}^2$. Note that the detection of the
neutrinoless double beta decay is practically the only way to determine
whether neutrinos are Dirac or Majorana particles. The discovery of such a
decay would also permit ones to determine the absolute scale of neutrino mass
using a measured value of $m_{\beta\beta}$.

We present below the expression for $m_{\beta\beta}$ through the masses of
neutrinos and the neutrino mixing parameters for the NH-case and IH-case,
respectively:
\begin{subequations}
\begin{align}
m_{\beta\beta}&=|c_{12}^2c_{13}^2m_1+s_{12}^2c_{13}^2m_2+s_{13}^2m_3
+\alpha^2m_{1'}|\,,\label{eq13a}\\
&\nonumber\\
m_{\beta\beta}&=|c_{12}^2c_{13}^2m_1+s_{12}^2c_{13}^2m_2+s_{13}^2m_3
+\gamma^2m_{3'}\nonumber\\
&+\delta^2m_{2'}+\delta^2m_{1'}|\,.\label{eq13b}
\end{align}
\label{eq13}
\end{subequations}
On the basis of the experimental data given in Eqs.~(\ref{eq4a})--(\ref{eq4e})
on the oscillation parameters of neutrino mixing and with the help of
numerical calculations for the given value of $m_{\beta\beta}$, it is possible
to estimate the absolute scale of neutrino mass spectra with normal and
inverse hierarchy. Indeed, with using the expression (\ref{eq13}) and the
experimental data (\ref{eq4}) one can determine the explicit dependences of
$m_{\beta\beta}$ on the smallest value $m_0$ from the neutrino masses, i.e.,
either on $m_1$ in the NH-case or on $m_3$ in the IH-case
(see Fig.~\ref{Fig1}, the left panel).

The same can be constructed for the values of other matrix elements $M_{ij}$,
as well as for the values of the neutrino mass observables $m_{\Sigma}$ and
$m_{\beta}$ (see Figs.~\ref{Fig1}--\ref{Fig2}). For example, the right panel
of Fig.~\ref{Fig1} shows the dependence of $m_{\beta}$ values on $m_0$, while
two panels, respectively, of Fig.~\ref{Fig2} shows the dependences of $M_{22}$
and $M_{33}$, on the values of $m_0$ simultaneously for both NH- and IH-cases.
As was noted above, the matrix elements of the mixing matrix are considered
as real in this paper, and, taking into account the results of Ref.~\cite{41},
it was adopted that $\delta_{CP}\simeq 0$ for the IH-case, while
$\delta_{CP}\simeq\pi$ for the NH-case.

It is possible to find and draw the dependences of the neutrino mass
observables on the greatest mass $m_*$ among the masses of all sterile
neutrinos (see Figs.~\ref{Fig3}(a)--\ref{Fig7}(a)). For comparison,
Figs.~\ref{Fig3}(b)--\ref{Fig7}(b) exhibit the behavior of these features
when there are no contributions of the sterile neutrinos. With the help of
these graphs, for example, different assumptions about the structure of the
mass matrix $M$ and the values of its matrix elements $M_{ij}$, which were
stated in a number of models \cite{42,43,44,45} can be tested. Making the
necessary calculations with the available experimental data and with minimal
arbitrariness, which are reduced to the choice of the values of $m_0$, it can
be possible to find a range of admissible values of $M_{ij}$, and thus to
check the validity of a model of the structure of the neutrino mass matrix.
For example, ranges of values calculated and shown in Fig.~\ref{Fig2} for the
diagonal matrix elements $M_{22}$ and $M_{33}$ confirm, first of all for the
NH-case, the applicability of commonly used approximate equality
$M_{22}=M_{33}$, which is a characteristic feature of the
$\mu$--$\tau$-symmetry in the neutrino sector \cite{32}.

To find the possible experimental effects associated with light sterile
neutrinos, it is of interest to consider the changes of the neutrino mass
observables due to the possible contributions of sterile neutrinos. For this
we choose as a determinative parameter the greatest mass $m_*$ of the sterile
neutrinos, which will be varied in the range from $0.1\,{\rm eV}$ to
$1.1\,{\rm eV}$, and then we obtain dependences of the neutrino mass
observables $m_{\Sigma}$, $m_{\beta}$ and $m_{\beta\beta}$ on $m_*$ taking
also into account the contributions of other sterile neutrinos. These
dependences are shown in Figs.~\ref{Fig3}(a)--\ref{Fig5}(a). Besides,
Figs.~\ref{Fig6}(a)--\ref{Fig7}(a) shows the dependences of the matrix
elements $M_{22}$ and $M_{33}$ on $m_*$. Although in this paper we consider
a sufficiently wide interval of possible values for the mass of the heavy
sterile neutrino, there are different estimates of the mass of sterile
neutrinos, which result to the values of the order of $1\,{\rm eV}$. For
example, two corresponding values, namely, $0.46\,{\rm eV}$ and
$0.96\,{\rm eV}$ were obtained in Ref.~\cite{46}.

As can be seen from these figures, the effect of sterile neutrinos in the
measurements of the mass observables $m_{\Sigma}$, $m_{\beta}$ and
$m_{\beta\beta}$ can not be detected at the sterile neutrinos masses
less than $0.5\,{\rm eV}$, if the condition of exceeding the background
level by more than $2\sigma$ is supposed to be fulfilled. On the
other hand, the recent results of measurements of the cosmological microwave
background lead to the fact that the sum of the masses of all types of
neutrinos does not exceed $0.5\,{\rm eV}$ \cite{14}. Thus, the search of the
effect of sterile neutrinos at the $(3\div 5)\sigma$ level should be performed
in the oscillation experiments, but not in the experiments for determination
of the absolute scale of neutrino masses.

The obtained ranges of possible values of $m_{\Sigma}$, $m_{\beta}$ and
$m_{\beta\beta}$ can be used in both the planning and the interpretation of
the results of the experiments to search for the neutrinoless double beta
decay, in the determination of the $\beta$-decay neutrino mass, as well as
in the determination of the cosmological sum of the absolute values of masses
of all neutrinos.

\section{Conclusion}%
\label{Sec7}%
Properties of neutrinos are very mysterious and intensive theoretical and
experimental studies to determine the nature and characteristics of these
elementary particles are required. Construction and development of adequate
phenomenological models of neutrinos, which can generalize the SM in the
neutrino sector, is one of the ways for correct interpretation and prediction
of the experimental results and successful searching the GUT. In the current,
the priority of the experimental and theoretical researches in the neutrino
physics is to verify the existence of light and heavy sterile neutrinos, to
determine their number, as well as to determine the absolute mass scale, both
for active and sterile neutrinos \cite{47}.

In the current work, the phenomenological $(3+1+2)$-model of neutrino with
three active and three sterile neutrinos, one of which is more massive and
the other two sterile neutrinos are much lighter is used to study the
properties of active and sterile neutrinos. The model allows reducing the
number of sterile neutrinos in case if the model-independent experimental
restrictions on their number will be established. On the basis of recent
experimental data, the allowable ranges of the neutrino mass observables
$ m_{\Sigma}$, $m_{\beta}$ and $m_{\beta\beta}$ have been calculated taking
into account the contributions of the sterile neutrinos. Using the estimates
of the masses of active neutrinos, the dependences of mass observables
$m_{\Sigma}$, $m_{\beta}$ and $m_{\beta\beta}$ on the mass of one of sterile
neutrino, which was varied in the range of possible values from
$0.1\,{\rm eV}$ to $1.1\,{\rm eV}$, were depicted. Currently, for experimental
determination of the neutrino mass observables $m_{\Sigma}$, $m_{\beta}$ and
$m_{\beta\beta}$ the numerous experiments are both carried out and planned on
searching for the neutrinoless double beta decay, defining the form of
beta-spectrum in the decay of tritium, as well as the cosmological
observations. To interpret and to predict the
results of these experiments, and also to explain the LSND/MiniBooNE, reactor
and gallium anomalies, the considered model of neutrinos and the
obtained values of the neutrino mass observables can be used.

\begin{acknowledgments}
The authors are grateful to Yu.~S. Lyutostansky, V.~P. Martemyanov,
M.~D. Skorokhvatov, S.~V. Sukhotin, S.~V. Semenov, D.~K. Nadezhin and
I.~V. Panov for useful discussions. This work was partially supported
by grants RFBR 11-02-00882-a and 12-02-12116-ofi-m.
\end{acknowledgments}

\end{document}